\documentclass[10pt]{iopart}
\usepackage{graphicx}
\usepackage{iopams}

\newcommand{\wb}{\omega_{\mathrm{b}}}
\newcommand{\ii}{\mathrm{i}}
\newcommand{\ee}{\varepsilon}

\newcommand{\sign}{\mathrm{sign}}

\newcommand{\fracc}[2]{\frac{\displaystyle #1}{\displaystyle #2}}

\newcommand{\doublefig}{0.9\textwidth}

\begin{document}

\title[Dark breathers stability]
{Dark breathers in Klein-Gordon lattices. Band analysis of their
stability properties.}

\author{A Alvarez \dag, JFR Archilla \ddag, J Cuevas \ddag \mbox{} and FR Romero \dag}

\address{Nonlinear Physics Group of the University of Sevilla.\\
\dag Facultad de F\'{\i}sica, P. 5 and \ddag Dep. F\'{\i}sica
Aplicada I en ETSI
Inform\'atica.\\
 Avda. Reina Mercedes s/n, 41012 Sevilla, Spain.}



\begin{abstract}
Discrete bright breathers are well known phenomena. They are
localized excitations that consist of a few excited oscillators in
a lattice and the rest of them having very small amplitude or
none. In this paper we are interested in the opposite kind of
localization, or discrete dark breathers, where most of the
oscillators are excited and one or a few units of them have very
small amplitude. We investigate using band analysis, Klein--Gordon
lattices at frequencies not close to the linear ones. Dark
breathers at low coupling are shown to be stable for Klein--Gordon
chains with soft on--site potentials and repulsive dispersive
interaction, and with hard on--site potentials and attractive
dispersive interactions. At higher coupling  dark breathers lose
their stability via subharmonic, harmonic or oscillatory
bifurcations, depending on the model. However, most of these
bifurcations are harmless in the sense that they preserve dark
localization. None of these bifurcations disappear when the system
is infinite. Dark breathers in dissipative systems are found to be
stable for both kinds of dispersive interaction.

\end{abstract}

\pacs{ 63.20.Pw,
       63.20.Ry,
       66.90.+r,
}



\maketitle

\section{Introduction}
\label{sec:intro}

It is well known that intrinsic localized modes (also called
\emph{discrete breathers}) are exact, periodic and localized
solutions that can be obtained in a large variety of nonlinear
discrete systems. They are becoming a new paradigm for
understanding many aspects of the behaviour of discrete systems
(for a review, see, e.g., \cite{A97,FW98}). Mackay and Aubry
\cite{MA94} proved analytically their existence and the conditions
for their stability \cite{A97}, under rather general hypotheses.
Since then, many accurate numerical methods have been used to
obtain breathers as exact numerical solutions up to machine
precision \cite{MA95}, which permits the analysis of breathers
properties. Thus, for a given model, it is possible to perform a
numerical study of the ranges of existence and stability in the
parameters space.

The term discrete breather is usually understood as a localized,
periodic solution in a discrete system, but with a small number of
excited oscillators. When only one oscillator has large amplitude
it is known as \emph{one--site breather}. When more than one
oscillator have large amplitude, the term \emph{multibreather} is
used.
>Hereafter, we will use the term \emph{bright breather} when there are one
or a few oscillators vibrating with large amplitude whereas the
rest of them oscillate with small amplitude.

However, localization can be manifested in a different way, which
consists of all the oscillators vibrating with large amplitude
except one or a few of them oscillating with very small amplitude.
The natural name for these entities is {\em dark breathers},
analogously to the well known term {\em dark solitons}. For the
nonlinear Schr\"odinger equations, which governs both nonlinear
optical modes in fibers and dilute Bose--Einstein condensates, two
different kinds of scalar soliton solutions, {\em bright and
dark}, are known \cite {ZS72,ZS73}. Thus, a dark soliton is a
solution which has a point with zero amplitude, that is, a soliton
defining the absence of matter or energy. Since then, many papers
have appeared referring to theoretical and experimental results
referred to this entities. The effects of discreteness on the
properties and propagation dynamics of dark solitons has been
analyzed in the discrete nonlinear Schr\"odinger equation (DNLS),
which is thought to be a good approximation for frequencies close
to the linear frequency \cite {KP92,KKC94,KT99,JK99,MJKA02}, the
last one, also with numerics on actual Klein--Gordon systems. It
is worth remarking that some examples of dark localization have
been observed experimentally \cite{DGG92}, and some structural
properties have been analyzed in \cite{DW95}.

In this paper, we perform an analysis of the existence and
stability of dark breathers for different models based on the
properties of the band structure of the Newton operator. We have
found that there exist stable dark breathers for a variety of
one-dimensional Klein-Gordon lattices. For soft on-site
potentials, dark breathers are stable only for repulsive
dispersive interaction and for hard on--site potentials the
stability founded is for attractive interaction. This results
agree with the ones that have been found for the DNLS
approximation in \cite{DDP97} as  a consequence of the
modulational instability of the constant amplitude background.

It is not clear in which physical systems dark breathers can play
a significant role. A possibility is  DNA, we conjecture that dark
breathers can occur in DNA chains at high temperature, close to
thermal denaturation \cite{PB89}. In this situation,  a great
number of molecules are vibrating with high amplitude, whereas a
few of them could be almost at rest.

This paper is organized as follows: In section \ref{sec:model}, we
describe the proposed Klein-Gordon models and the resulting
evolution equations. In Section \ref{sec:darkexistence}, we expose
the tools for calculating dark breathers, and we explicitly show
that the theorem of existence of Mackay and Aubry gives an
affirmative answer to the question of existence of dark breathers.
In section \ref{sec:darkstability}, we expose the method for the
analysis of the linear stability of breather solutions  using both
the Floquet multipliers and Aubry band's theory. In Section
\ref{sec:soft}, we investigate the stability of dark breathers for
chains with soft on-site potential and attractive interactions
between the particles. We show that it is not possible to obtain
stable dark breathers for every value of the coupling parameter.
This negative result suggested us to consider the study of chains
with repulsive interactions between  particles. Our results
confirm that in this case there exist stable dark breathers up to
significant values of the coupling. In Section
\ref{sec:hardbreathers}, lattices with hard on-site potentials are
considered. For these cases, dark breathers are stable provided
that particles interact through an attractive potential. Dark
breathers become unstable through subharmonic, harmonic and
oscillatory bifurcations depending on the type of the on--site
potential. In section \ref{sec:finite}, we show that both
subharmonic and oscillatory bifurcations do not disappear in
infinite systems. The paper concludes with a short summary of the
main results and some prospective in Section
\ref{sec:conclusions}.

\section{Models}
\label{sec:model}

We study one--dimensional, anharmonic, Hamiltonian lattices of the
Klein--Gordon type. The Hamiltonian is given by
\begin{equation}
  H=\sum_{n} (\frac{1}{2}\,\dot{u}_n^2+V(u_n))+ \varepsilon\, W(u)
\label{eq:ham1}
\end{equation}
where $u_n$ are the coordinates of the oscillators referred to
their equilibrium positions; $V(u_n)$ represents the on--site
potential; $u$ represents the set of variables $\{u_n\}$; and
$\varepsilon W(u)$ represents the coupling potential, with
$\varepsilon$ being a parameter that describes the strength of the
coupling. We  suppose initially that $\varepsilon$ is positive and
$W(u)$ is given by
\begin{equation}
W(u)=\frac{1}{2}\,\sum_{n}(u_{n+1}-u_n)^2.
\end{equation}
This interaction is attractive because a nonzero value of a
variable tends to increase the values of the neighbouring
variables with the same sign. The on--site potential is given by
\begin{equation}
V(u_n)=\frac{1}{2}\,\omega_0^2\;u_n^2+\,\phi(u_n)
\end{equation}
with $\phi(u_n)$ being the anharmonic part of the potential. The
variables are scaled so that all the particles in the lattice have
mass unity and the linear frequency $\omega_0=1$. The dynamical
equations for this system are
\begin{equation}
\ddot{u}_n+\omega_0^2u_n+\phi'(u_n)+\varepsilon(2u_n-u_{n-1}-u_{n+1})=0.
 \label{eq:ecmov1}
\end{equation}
These equations do not have analytical solutions and must be
solved numerically. The solutions depend obviously on the chosen
potentials $V(u_n)$ and $W(u)$. We will analyze the system for
several $V(u_n)$ and coupling interactions $W(u)$. These models
appear in many physical systems, a known example, with a suitable
on-site potential $V(u_n)$ is the Peyrard--Bishop model for DNA
\cite{PB89}, where the variables $u_n$ represent the stretching of
the base pairs.

\section{Dark breathers existence}
\label{sec:darkexistence}

We look for spatially localized, time--reversible and
time--periodic solutions of equations~(\ref{eq:ecmov1}) with a
given frequency $\omega_b$ and a continuous second derivative.
Therefore, the functions $u_n(t)$ can be obtained up to machine
precision  by truncated Fourier series of the form
\begin{equation}
\label{eq:fourier}
 u_n(t)=z_0+\sum_{k=1}^{k=k_m}2 z_n^k\cos(k \wb t).
\end{equation}

 We distinguish three types of solutions of the isolated
oscillators that can be coded in the following way: $\sigma_n=0$
for an oscillator at rest $(u_n(t)=0, \; \forall t)$;
$\sigma_n=+1$ for an excited oscillator with frequency $\wb$ and
$u_n(0)>0$; finally, $\sigma_n=-1$ for an excited oscillator with
frequency $\wb$ and $u_n(0)<0$. Time--reversible solutions of the
whole system at $\varepsilon=0$ (anticontinuous limit), can be
referred to by a coding sequence $\sigma=\{\sigma_n\}$. Therefore,
$\sigma=\{0,...,0,1,0,...,0\}$ corresponds to a one--site
breather. Other codes can be $\sigma=\{0,...,0,1,1,0,...,0\}$ for
a symmetric two--site breather, and $\sigma=\{1,...,1,0,1,...,1\}$
for a one--site dark breather (this case correspond to the
background in phase).

The method for calculating dark breathers is based on the general
methods for obtaining breathers
\cite{MA95,Fla95a,Fla95b,MA96,MAF98}.

The existence theorem by MacKay and Aubry~\cite{MA94} establishes
that every solution at the anticontinuous limit, corresponding to
a code sequence can be continued up to a certain value of the
coupling parameter $\varepsilon_c\neq 0$, as long as the following
two hypotheses are fulfilled:
\begin{itemize}
 \item The orbits of the uncoupled excited oscillators with the chosen
 frequency have to be such that
 $\frac{\partial\wb}{\partial I}\neq0$, where $I=\int p d q$ is
 the action variable of the oscillator. That is, the oscillator is truly
 nonlinear at that frequency.
 \item The frequency of the orbit must be such that
 $p\omega_b\neq\omega_0$ for any integer $p$.
 That is, none of the breather harmonics coincide with the
 linear frequency $\omega_0$.
 \end{itemize}

Therefore, this theorem gives an immediate answer to the question
of the existence of dark breathers as they are obtained by
continuation of the configuration mentioned above
$\sigma=\{1,...,1,0,1,...,1\}$ . However, the theorem does not
give an estimate for the value of the coupling $\varepsilon_c$
where the dark breather ceases to exist. Dark breathers have to be
calculated numerically for each value of the coupling parameter,
also it is worth investigating whether dark breathers are stable
or not, and if they are stable, up to which value of the coupling
parameter.

\section{Dark breathers stability}
\label{sec:darkstability}

The stability analysis of a given breather solution can be
performed numerically \cite{A97,MAF98,MS98,M97}. The linearized
equations corresponding to perturbations of this solution are
\begin{equation}
  \ddot{\xi}_n\,+\,\omega_0^2\,\xi_n+\phi''(u_n)\,\xi_n + \varepsilon
  (2\xi_n-\xi_{n-1}-\xi_{n+1})  = 0
\label{eq:Newtonop1}
\end{equation}
where $\xi=\{\xi_n(t)\}$ represents a small perturbation of the
solution of the dynamical equations, $u(t)=\{u_n(t)\}$. The linear
stability of these solutions can be studied by finding the
eigenvalues of the Floquet matrix $\mathcal{F}_0$, called Floquet
multipliers. The Floquet matrix transforms the column matrix with
elements given by $\xi_n(0)$ and $\pi(0)\equiv \dot{\xi}_n(0)$
into the corresponding column matrix with elements $\xi_n(T_b)$
and $\pi(T_b)\equiv \dot{\xi}_n(T_b)$ for $n=1...\, m$ and
$T_b=2\pi/\wb$, that is
\begin{equation}
  \left( \begin{array}{c}\{ \xi_n(T_b)\} \\ \{\pi_n(T_b)\} \end{array}
  \right) = \mathcal{F}_0 \left( \begin{array}{c}\{ \xi_n(0) \}\\ \{ \pi_n(0)\}
    \end{array} \right)
  \label{eq:monodromy}
\end{equation}
The Floquet matrix $\mathcal{F}_0$ can be obtained numerically by
integrating equations~(\ref{eq:Newtonop1})  a breather period. In
order to get accurate results ~\cite{SC94}, we have used a
symplectic integrator. Equation~(\ref{eq:Newtonop1}) can be
written as an eigenvalue equation
\begin{equation}
(\mathcal{N}(u(t),\varepsilon)\cdot \xi\,)_n\,=\,E\,\xi_n
 \label{eq:Newton}
\end{equation}
where $\mathcal{N}(u(t),\varepsilon)\}$ is called the Newton
operator. The solutions of equation~(\ref{eq:Newtonop1}) can be
described as the eigenfunctions of $\mathcal{N}$ for $E=0$. The
fact that the linearized system is Hamiltonian and real implies
that the Floquet operator is a real and symplectic operator. The
consequence is that if $\lambda$ is an eigenvalue, then
$1/\lambda$, $\lambda^{*}$ and $1/\lambda^{*}$ are also
eigenvalues, and therefore a necessary condition for linear
stability is that every eigenvalue has modulus one, that is, they
are located at the unit circle in the complex plane. Besides,
there is always a double eigenvalue at $1+0i$ from the fact that
the derivative $\{\dot{u}_n(t)\}$ is always a solution of
(\ref{eq:Newton}) with $E=0$.

The  Floquet multipliers at the anticontinuous limit
$(\varepsilon=0)$ can be easily obtained for bright and dark
breathers. If an oscillator at rest is considered, equation
(\ref{eq:Newtonop1}) becomes
\begin{equation}
  \ddot{\xi}_n\,+\,\omega_0^2\,\xi_n= 0 \label{eq:Newtonop1rest}
\end{equation}
with solution $\xi(t)=\xi_0\,e^{i\omega_0t}$, and therefore, the
corresponding eigenvalue of $\mathcal{F}_0$ is $\lambda=\exp(\ii
2\pi\omega_0/\omega_b)$. If an oscillator is excited, equation
(\ref{eq:Newtonop1}) becomes
\begin{equation}
  \ddot{\xi}_n\,+\,\omega_0^2\,\xi_n+\,\phi''(u_n)\,\xi_n= 0.
\label{eq:Newtonop1exc}
\end{equation}
This equation has $\dot{u}_n(t)$ as a solution, which is periodic
and therefore with Floquet multiplier $\lambda=1$. Thus, for a
bright breather, we have, taking into account their multiplicity,
$2(N-1)$ eigenvalues corresponding to the rest oscillators at
$\exp(\pm \ii 2\pi\omega_0/\omega_b)$ and two at $+1$
corresponding to the excited one. For a dark breather at the
anticontinuous limit there are $2(N-1)$ eigenvalues at $+1$, and a
couple of conjugate eigenvalues at $\exp(\pm \ii
2\pi\omega_0/\omega_b)$.

\begin{figure}
  \begin{center}
    \includegraphics[width=0.9\textwidth]{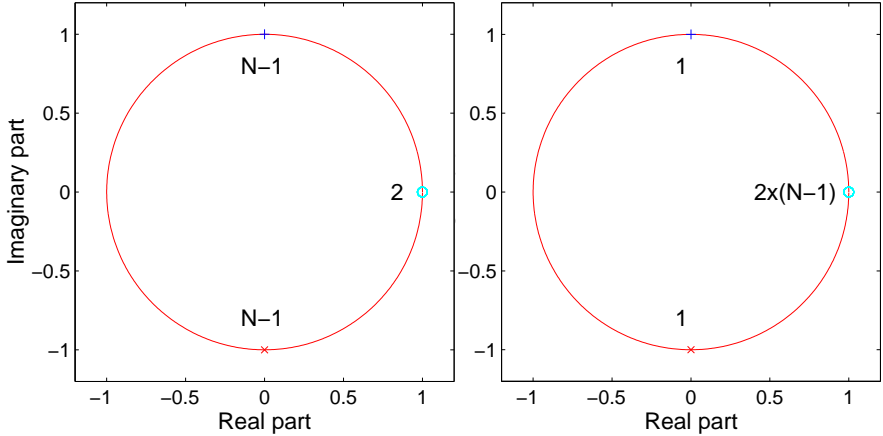}
  \end{center}
  \caption{Floquet multipliers at zero coupling for: (left) a bright
  breather and (right) a dark breather with the
  numbers of identical multipliers.}
  \label{fig:cire0}
\end{figure}

Figure~\ref{fig:cire0} shows the Floquet multipliers for both a
bright and a dark breather at the anti-continuous limit $\ee=0$.
When the coupling $\varepsilon$ is switched on, these eigenvalues
move on the complex plane as continuous functions of
$\varepsilon$, and an instability can be produced only in three
different ways~\cite{A97}~:\,(a)~a couple of conjugate eigenvalues
reaches the value $1+0\ii$ $(\theta=0)$\, and leaves the unit
circle along the real line (harmonic bifurcation); (b) a couple of
conjugate eigenvalues reaches $-1$ $(\theta=\pm\pi)$ and leaves
the unit circle along the real line (subharmonic bifurcation); (c)
two pairs of conjugate eigenvalues collide at two conjugate points
on the unit circle and leave it (Krein crunch or oscillatory
bifurcation). It must be remembered that  a bifurcation involving
two eigenvalues with the same sign of the Krein signature
$\kappa(\theta)=\sign(\ii(\dot{\xi}\cdot\xi^*-\dot{\xi}^*\cdot\xi))$
is not possible~\cite{A97}.

The basic features of the Floquet multipliers at the
anti-continuous limit for bright and dark breathers are the
following:

\begin{itemize}

\item
For bright breathers, there are $N-1$ Floquet multipliers
corresponding to the oscillators at rest. They are degenerated at
$\theta=\pm2\pi\omega_0/\wb)$. The eigenvalues at $0<\theta<\pi$
have $\kappa>0$ while the ones at $-\pi<\theta<0$ have $\kappa<0$.
If $\theta=0$ or $\pi$ $\kappa=0$. In addition, there are two
eigenvalues at $1+0\ii$, corresponding to the excited oscillator.

\item
For dark breathers, there are two eigenvalues at $\theta=\pm2\pi
\omega_0/\wb$ corresponding to the oscillator at rest, and there
are $2(N-1)$ eigenvalues at $1+0\ii$ corresponding to the excited
oscillators.

\end{itemize}

When the coupling is switched on, the evolution of the Floquet
multipliers for bright breathers is rather different from the case
of dark breathers:

\begin{itemize}

\item
For a bright breather, the Floquet eigenvalues corresponding to
the oscillators at rest lose their degeneracy and expand on two
bands of eigenvalues, called the phonon bands. Their corresponding
eigenmodes are extended. These two bands move on the unit circle
and eventually cross each other. In this case, eigenvalues of
different Krein signature can collide and abandon the unit circle
through subharmonic or oscillatory bifurcations. In addition, some
eigenvalues can abandon the bands and become localized
\cite{BKM98}. A pair of complex conjugate eigenvalues can collide
at $1+0\ii$ leading to a harmonic bifurcation or collide with the
phonon band through a oscillatory bifurcation.

\item
For a dark breather, the eigenvalues corresponding to the excited
oscillators, with extended phonon eigenmodes, can either depart
from the unit circle along the real axis (harmonic bifurcation) or
move along the unit circle. In the last case, they can collide
with the eigenvalue corresponding to the rest oscillator (with
localized eigenmode) through a Krein crunch. Eventually, if the
on-site potential is non-symmetric, the eigenvalues collide at
$-1+0\ii$, leading to a cascade of subharmonic bifurcations.

\end{itemize}

The study of breather stability can be complemented by means of
Aubry's band theory~\cite{A97}. It consists in studying the
linearized system (\ref{eq:Newton}) for $E\neq0$, with the
corresponding family of Floquet operators $\mathcal{F}_E$. For
each operator $\mathcal{F}_E$ there are $2\times N$ Floquet
multipliers. A Floquet multiplier can be written as
$\lambda=\exp(\ii \theta)$. $\theta$ is called the Floquet
argument. If $\theta$ is real then $|\exp(\ii \theta)|=1$ and the
corresponding eigenfunction of (\ref{eq:Newton}) is bounded and
corresponds to a stability mode; if $\theta$ is complex, it
corresponds to an instability mode. The set of points
$(\theta,E)$, with $\theta$ real, has a band structure. The
breather is stable if there are $2\times N$ band intersections
(including tangent points with their multiplicity) with the axis
$E=0$. The bands are reduced to the first Brillouin zone
$(-\pi,\pi]$ and are symmetric with respect to the axis
$\theta=0$. The fact that $\mathcal{F}_0$ has always a double $+1$
eigenvalue corresponding to the phase mode $\dot{u}(t)$ manifests
as a band which is tangent to the $E=0$ axis.

For the uncoupled system, the structure of the stable and unstable
bands is completely explained by the theory (although it has to be
calculated numerically). For the coupled system, such structure is
expected to change in a continuous way in terms of the parameter
$\varepsilon$. We will use band theory, to predict the evolution
of the  eigenvalues of $\mathcal{F}_0$, with a model parameter is
changed.

\begin{figure}
  \begin{center}
    \includegraphics[width=0.9\textwidth]{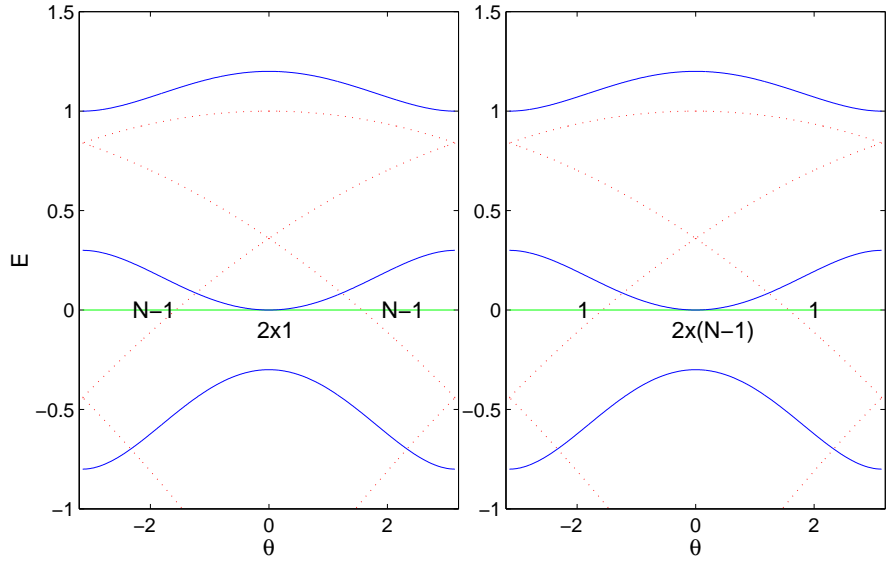}
  \end{center}
  \caption{Band scheme for bright (left) and dark (right)
  breathers with soft potential. The continuous lines correspond to the
  excited oscillators and are numbered downwards starting from zero.
  The dotted lines correspond to the oscillator at rest.}
  \label{fig:BandschemeBD}
\end{figure}

\begin{figure}
  \begin{center}
    \includegraphics[width=0.75\textwidth]{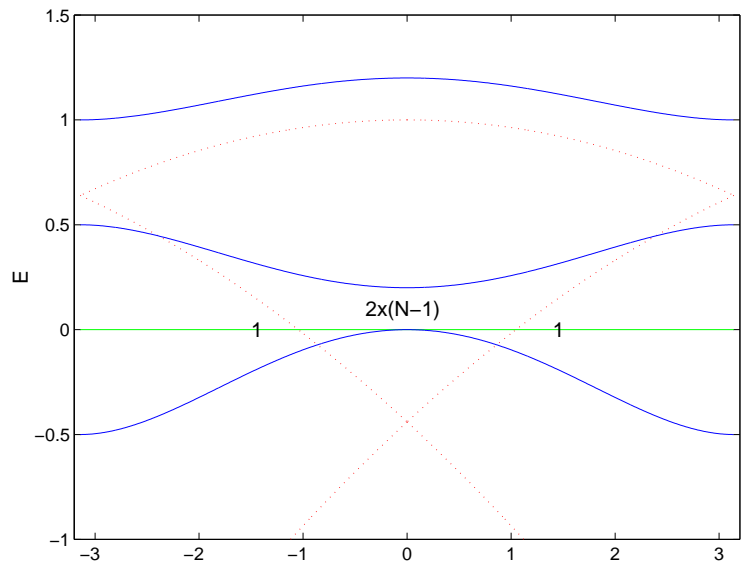}
  \end{center}
  \caption{Bands scheme for a dark breather with hard potential
  at zero coupling. The continuous lines correspond to the
  excited oscillators and are numbered downwards starting from zero.
  The dotted lines correspond to the oscillator at rest.}
  \label{fig:darkbands0}
\end{figure}

At zero coupling, the bands corresponding to the oscillators at
rest can be analytically calculated from the equation:
\begin{equation}
 \ddot{\xi}_n\,+\,\omega_0^2\,\xi_n= E\,\xi_n,
 \label{eq:B-anticon}
\end{equation}
they are given by $E=\omega_0^2-\omega_b^2 \theta^2/(2\pi)^2$. The
bands corresponding to the $N-1$ excited oscillators are a
deformation of the band corresponding to the oscillator at rest
(which will be called \emph{rest bands}) and one of them must be
tangent to $E=0$ axis at $(\theta,E)=(0,0)$ \cite{A97}. The bands
are bounded from above and numbered starting from the top, being
the 0th band the first one. If the on--site potential is soft the
1st band will be tangent to the $E=0$ axis at $(0,0)$ with
positive curvature. If the on--site potential is hard, the tangent
band at $(0,0)$ will be the 2nd one, and will have negative
curvature \cite{M97}. At the anticontinuous limit the band scheme
of a bright and a dark breather are similar, differing only in the
number of bands that corresponds to the oscillators at rest or to
the excited oscillators (\emph{excited bands}) as it is shown in
Figures \ref{fig:BandschemeBD} and \ref{fig:darkbands0}.

The bright breather has only one band tangent to the axis $E=0$
and that band cannot leave this position because, when the
coupling is switched on, there must be a tangent band
corresponding to the phase mode. The situation is totally
different for a dark breather: there are $N-1$ bands tangent to
the $E=0$ axis and $N-2$ of them can move without any restriction.
If some of them move upwards, the intersection points disappear,
which implies that some Floquet arguments become complex or,
equivalently, that some Floquet multipliers abandon the unit
circle and the breather becomes unstable.

\section{Dark breathers with soft on--site potentials}
\label{sec:soft}

\begin{figure}
  \begin{center}
    \includegraphics[width=0.9\textwidth]{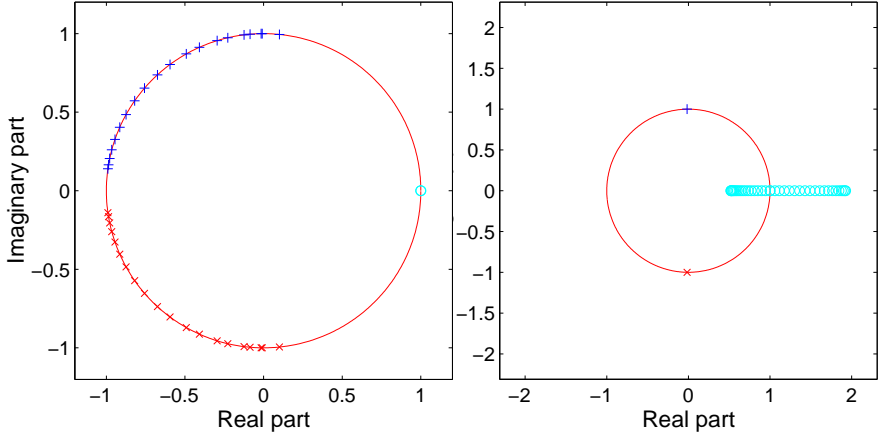}
  \end{center}
  \caption{Evolution of the Floquet multipliers with cubic
  on--site potential and attractive interaction when the coupling
  is switched on for: (left) the bright breather at $\varepsilon=0.1$,
  which is linearly stable, although reaching a possible bifurcation
  at $-1$; (right) the dark breather at a much smaller coupling parameter
  $\varepsilon=0.004$. The breather frequency is $\wb=0.8$.}
  \label{fig:cire01e004}
\end{figure}

Let us consider a model with a cubic on--site potential given by
\begin{equation}
V(u_n)=\frac{1}{2}\,\omega_0^2\;u_n^2-\,\frac{1}{3}u_{n}^3,
\end{equation}
that is, $\phi'(u_n)=-u_n^2$ in the dynamical equations
(\ref{eq:ecmov1}). Figure \ref{fig:cire01e004} represents the
Floquet multipliers for $\varepsilon\neq0$. The left side of the
figure shows the Floquet multipliers of a stable bright breather
for a coupling $\ee=0.1$. However, as  can be seen at the right
side of the figure, the dark breather experiences a multiple
harmonic bifurcation as soon as the coupling is switched on.
Therefore, for a lattice with a cubic on--site potential and
attractive interaction, dark breathers exist but they are unstable
for every value of the coupling.

\begin{figure}
  \begin{center}
    \includegraphics[width=0.75\textwidth]{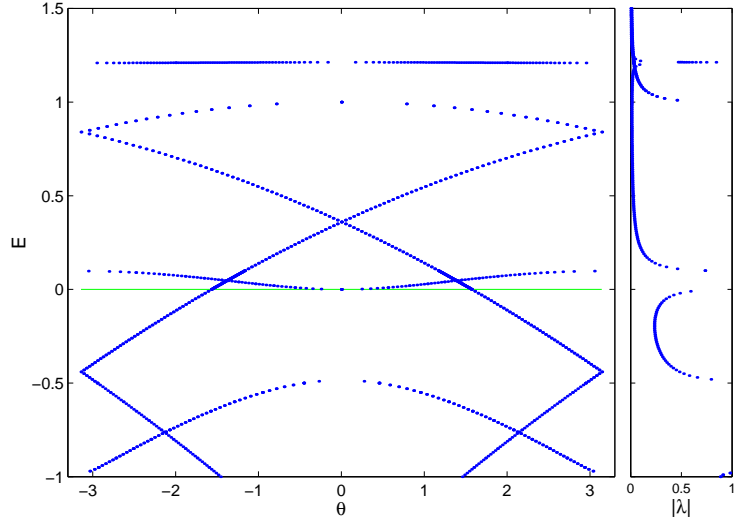}
   \end{center}
  \caption{Band structure at zero coupling for a cubic on--site
  potential. The right part of the figure plots the moduli of the unstable Floquet multipliers
  that are smaller than $1$, being their inverses the unstable
  multipliers. Frequency $\wb=0.8$}.
  \label{fig:bandcube0}
\end{figure}

 This problem can be investigated by means of Aubry's band
analysis. It has given us  an explanation for the previous
behaviour and also a guide for modifying the model in order to
obtain stable dark breathers. The bands at zero coupling can be
seen in Figure~\ref{fig:bandcube0}.

\begin{figure}
  \begin{center}
    \includegraphics[width=0.75\textwidth]{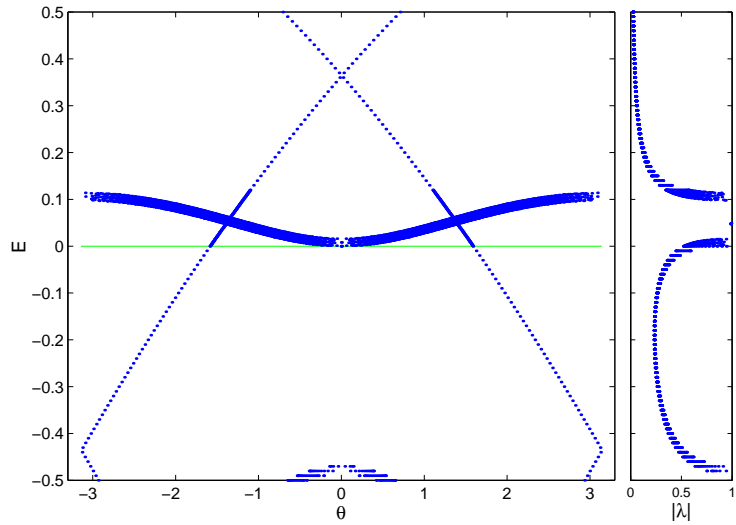}
   \end{center}
  \caption{Band structure of a dark breather with cubic on--site
   potential and attractive interaction at very
   low coupling ($\varepsilon=0.004$). Frequency $\wb=0.8$.}
  \label{fig:bandcubste004}
\end{figure}

\begin{figure}
  \begin{center}
    \includegraphics[width=0.75\textwidth]{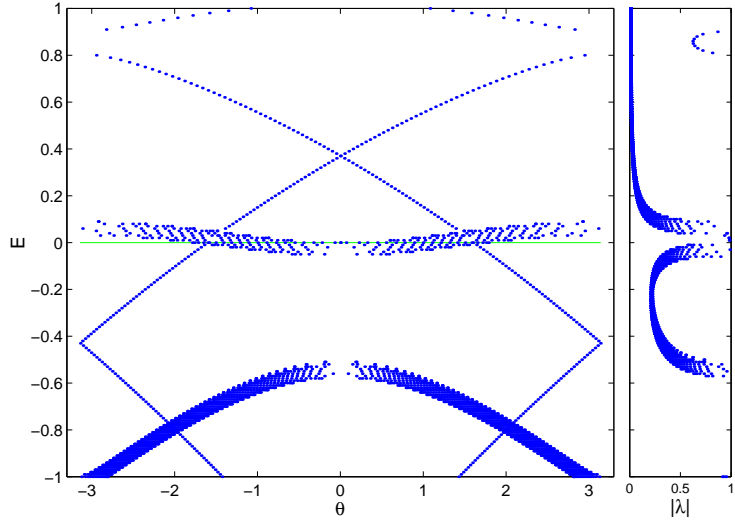}
   \end{center}
  \caption{Band structure of a dark breather with  cubic on--site
   potential and  repulsive interaction at $\varepsilon=0.015$.
   Frequency $\wb=0.8$.}
  \label{fig:Bcubdip015}
\end{figure}

As it is explained above and shown in Figure \ref{fig:cire01e004},
the cubic dark breather becomes unstable for any attractive
coupling $\varepsilon>0$ through harmonic bifurcations. This is
easily understood in terms of the band structure: $N-2$ tangent
bands move upwards, which is mathematically demonstrated in
Ref.~\cite{ACSA02}. They lose the tangent points with the $E=0$
axis as Figure \ref{fig:bandcubste004} shows. Therefore, in order
that the breather can be stable all the excited bands except one
have to move downwards, transforming the points tangent to the
$E=0$ axis into intersection points. A straightforward alternative
is to change the sign of the parameter $\ee$ in (\ref{eq:ham1}).
This is equivalent to using a dipole-dipole coupling potential
$W(u)=\sum_n\,u_{n+1}\,u_n$, i.e., the Hamiltonian can be written
as:
\begin{equation}
 H=\sum_n \left(\fracc{1}{2}\dot{u}_n^2
+\fracc{1}{2}\omega_0^2 u_n^2
 -\fracc{1}{3}u_n^3+\fracc{1}{2}\varepsilon(u_{n+1}\,u_n) \right)
 \end{equation}
with $\ee>0$. This repulsive interaction has been used recently to
model  DNA--related models
\cite{CGJ98,GMC00,ACG01,CGM01,CAGR02,CPAR02}.
  The dynamical equations with  repulsive
interaction and cubic on--site potential become:
\begin{equation}
\ddot{u}_n+\omega_0^2u_n-u_n^2+\varepsilon(u_{n-1}+u_{n+1})=0,
\label{eq:ecmov2}
\end{equation}
and the linear stability equations become:
\begin{equation}
  \ddot{\xi}_n\,+\,\omega_0^2\,\xi_n\,-\,2u_n\,\xi_n + \varepsilon
  (\xi_{n-1}+\xi_{n+1}) = 0.
\label{eq:Newtonop2}
\end{equation}
Figure \ref{fig:Bcubdip015} displays the band structure at
$\varepsilon=0.015$ for this system. The $N-2$ bands that are
allowed to move will perform a downwards movement and therefore
the breather is stable.

\begin{figure}
  \begin{center}
    \includegraphics[width=0.75\textwidth]{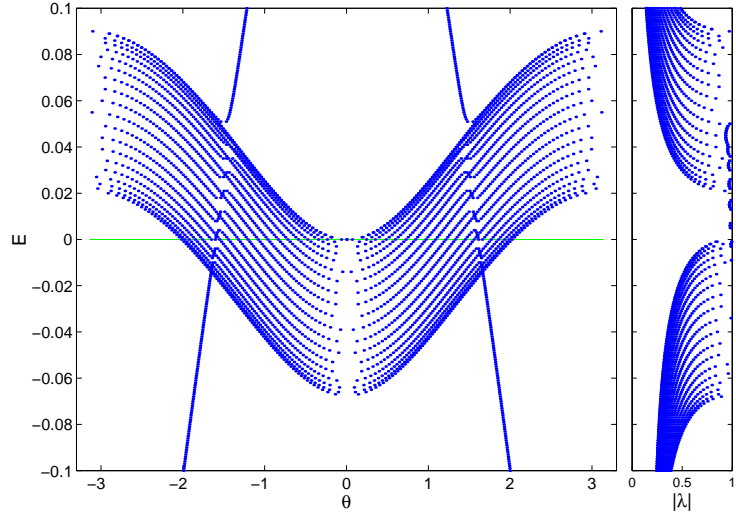}
   \end{center}
  \caption{Band structure of a dark breather with cubic on--site
   potential and repulsive interaction at $\varepsilon=0.017$ and
   $\wb=0.8$. Note the ``wiggles'' that appear when the
   excited bands and the rest band mix.
   A further increase of the coupling parameter $\ee$ will make
   the excited bands fall below $E=0$ producing subharmonic bifurcations.}
  \label{fig:Bcubdipe017}
\end{figure}

A further increase of $\varepsilon$ leads to Krein crunches. They
are caused by the mixing of the rest bands and the excited bands,
that produces ``wiggles'' in the excited bands and gaps appears
between them \cite{MA98}. When these bands move downwards and the
``wiggles'' cross the $E=0$ axis, intersection points are lost but
recovered when the coupling increases. It manifests as the
appearance of ``instability bubbles''. These instabilities depend
on the size of the system, as will be shown in section
\ref{sec:finite}. The system will eventually become unstable
through a cascade of subharmonic bifurcations when the lower
excited band goes below the $E=0$ axis, losing the intersection
points at $\theta=\pm\pi$ (see figure \ref{fig:Bcubdipe017}).

\begin{figure}
  \begin{center}
    \includegraphics[width=0.75\textwidth]{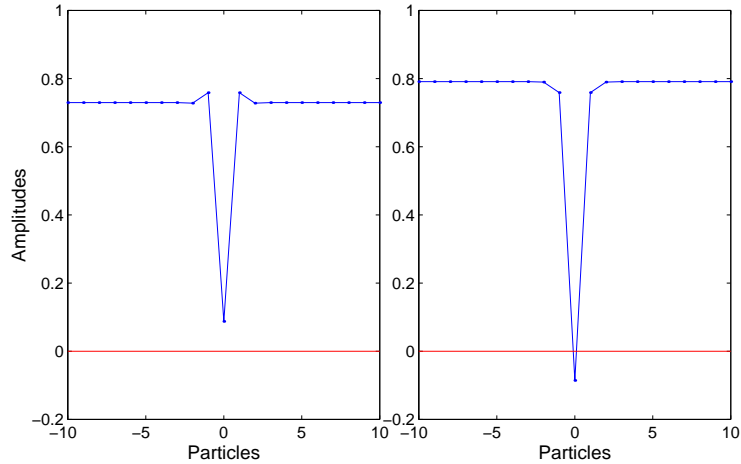}
   \end{center}
  \caption{Dark breathers profile for a cubic potential with
  attractive interaction (left) and with repulsive
  interaction(right), at $\varepsilon=0.023$ . The
  last one is stable.}
  \label{fig:twodarks}
\end{figure}

We have obtained dark breathers for systems with cubic on--site
potential and  attractive or  repulsive interaction even though
dark breathers with attractive coupling are not stable. Figure
\ref{fig:twodarks} shows two different examples of dark breathers
for this two types of coupling. Notice that the oscillator with
small amplitude is in phase with its neighbours when the system is
unstable (left) while is in anti--phase when it is stable (right).

\begin{figure}
  \begin{center}
    \includegraphics[width=0.75\textwidth]{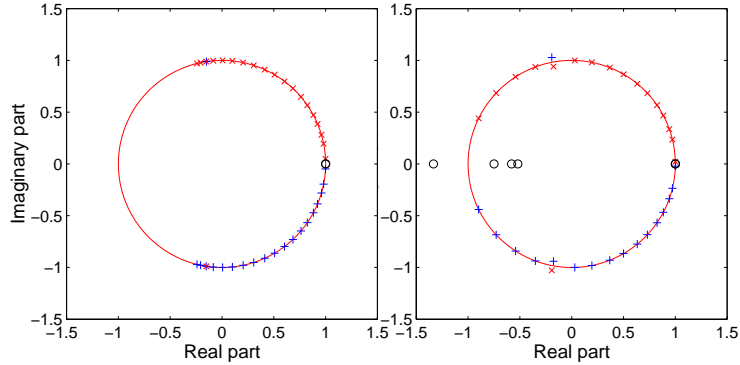}
   \end{center}
  \caption{Evolution of the Floquet eigenvalues for a Morse
  on--site potential and repulsive coupling. Left: $\varepsilon=0.024$,
  the system is still stable. Right: $\varepsilon=0.035$,
  the system becomes unstable due to subharmonic bifurcation and
  Krein crunches.}
  \label{fig:twocirclesMorse}
\end{figure}

 This type of behaviour is general for other soft on--site
potentials as it can be obtained using, for example, the Morse
potential given by
\begin{equation}
V(u_n)=D\,(\exp(-bu_n)-1)^2.
\end{equation}
For this system, dark breathers maintain their stability until the
coupling parameter reaches the value $\varepsilon=0.024$. From
this value Krein bifurcations appear. These bifurcation are
harmless in the sense that they preserve the dark localization,
but the breather becomes quasiperiodic with a superimposed
frequency to the breather one.
 For $\varepsilon>0.033$,
the system becomes unstable due to subharmonic bifurcations.
Figure~\ref{fig:twocirclesMorse} shows this behaviour in terms of
the corresponding Floquet multipliers for two different values of
the coupling. With this potential, it has been described
\cite{MJKA02} that for values of the $\wb$ close to the linear
frequency $\omega_0$ (in fact, slightly above it), oscillatory
instabilities can bring about the movement of  dark breathers,
when the corresponding eigenvalue is asymmetric and localized. It
does not happen for frequencies far enough of $\omega_0$ as the
ones of the order we present here.

\begin{figure}
  \begin{center}
    \includegraphics[width=0.75\textwidth]{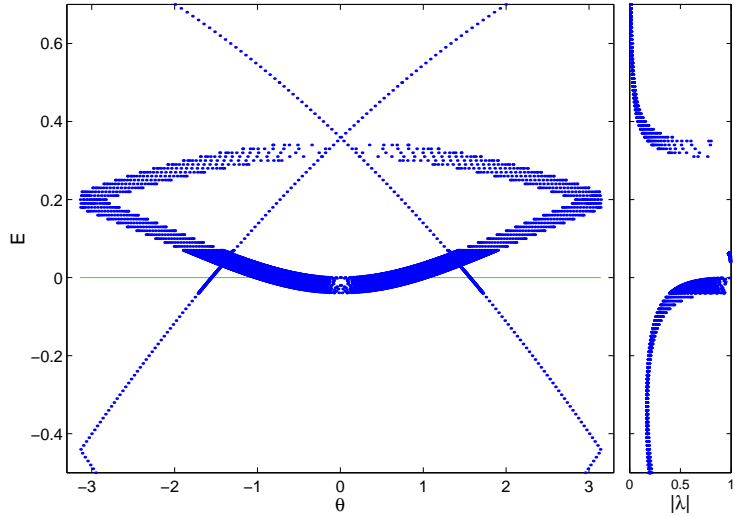}
  \end{center}
  \caption{Band structure for a dark breather with
   quartic soft on--site potential and repulsive coupling.
   Note the absence of gaps at $\theta=\pm\pi$ which enlarges
   the stability range. Parameters: $\varepsilon=0.018$ and $\wb=1.2$.}
  \label{fig:Bquartsofdip018}
\end{figure}

An interesting variation of this scheme occurs if we consider
symmetric on--site potentials as the quartic soft potential given
by
\begin{equation}
V(u_n)=\frac{1}{2}\,\omega_0^2\;u_n^2-\,\frac{1}{4}u_{n}^4.
\end{equation}
In this case the potential is symmetric and no subharmonic
bifurcations at $\lambda=\pm1$ occur because the bands are gapless
at $\theta=\pm\pi$. This characteristic enlarges the stability
range (see Figure \ref{fig:Bquartsofdip018}). The system
eventually becomes unstable (apart from the reentrant
instabilities due to Krein crunches) through harmonic
bifurcations.

\section{Dark breathers with hard on--site potentials}
\label{sec:hardbreathers}

The band scheme of a system with hard on--site potential at
$\varepsilon=0$ is shown in Figure \ref{fig:darkbands0}. There are
$N-1$ bands tangent to the axis $E=0$. It is clear that the
breather will remain stable provided that the tangent bands keep
the intersection points when the degeneracy is raised with
$\ee\neq0$. In this system it means that the $N-2$ bands that can
move will perform an upwards movement. Therefore, the dark
breather will be stable with an attractive coupling potential. We
have used for the numerics the quartic hard on--site potential,
that is
\begin{equation}
V(u_n)=\frac{1}{2}\,\omega_0^2\;u_n^2+\,\frac{1}{4}u_{n}^4.
\end{equation}
The Hamiltonian is then
\begin{equation}
 H=\sum_n \left(\fracc{1}{2}\dot{u}_n^2
+\fracc{1}{2}\omega_0^2 u_n^2
 +\fracc{1}{3}u_n^3+\fracc{1}{2}\varepsilon(u_{n+1}-\,u_n)^{2} \right)
 \end{equation}

\begin{figure}
  \begin{center}
    \includegraphics[width=0.75\textwidth]{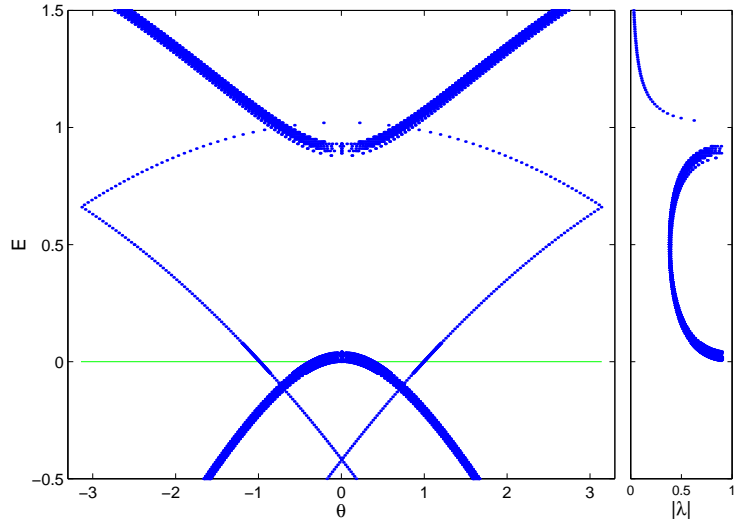}
   \end{center}
  \caption{Band structure for a quartic hard on--site potential
   with attractive interaction at $\varepsilon=0.01$ and $\wb=1.2$.}
  \label{fig:Bquarhardsta01}
\end{figure}

\begin{figure}
  \begin{center}
    \includegraphics[width=\doublefig]{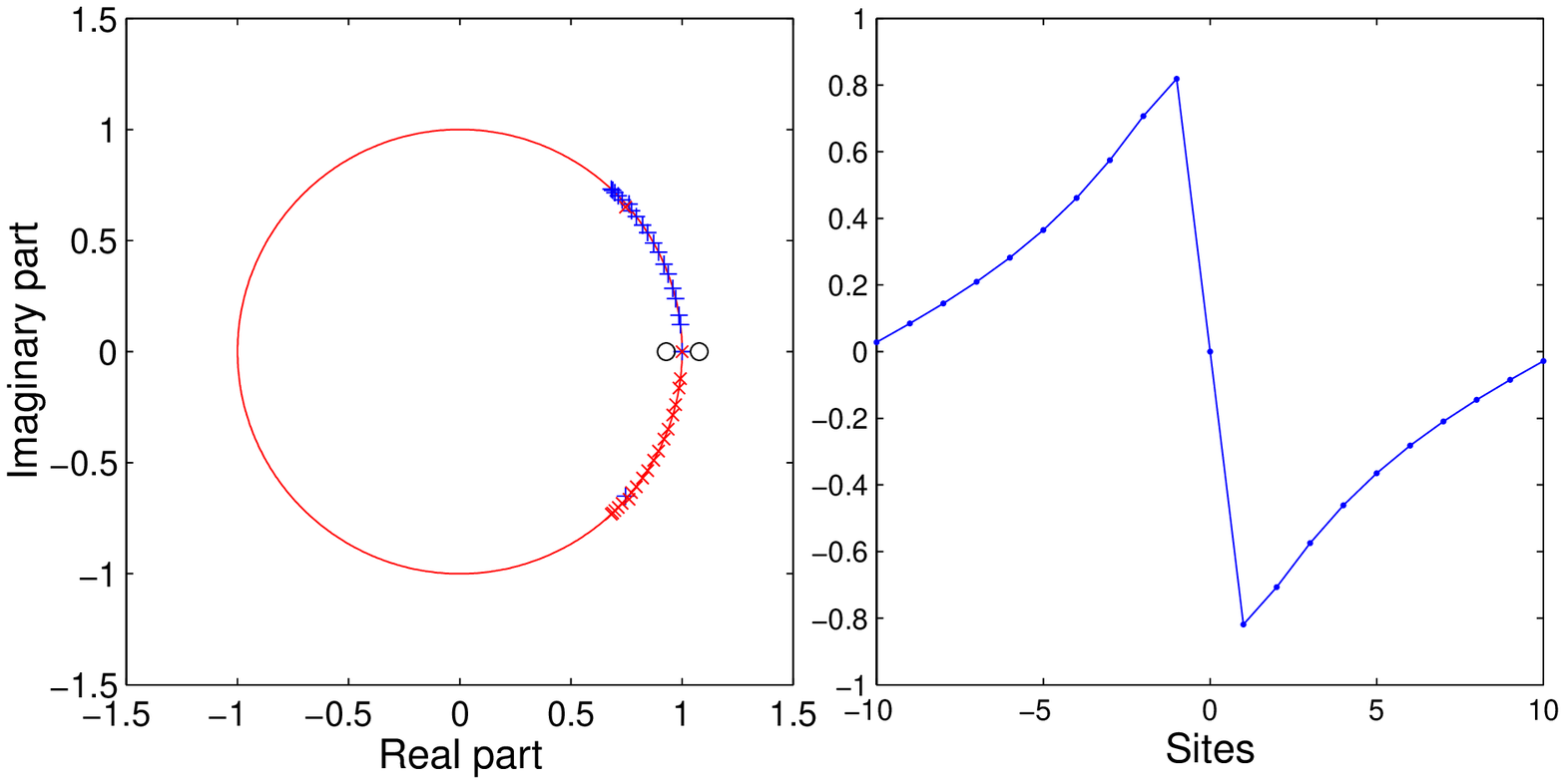}
  \end{center}
  \caption{Left: Floquet multipliers with a quartic hard on--site potential
   and attractive interaction at $\varepsilon=0.041$ and $\wb=1.2$. A
   harmonic and a small Krein instabilities appear.Right: the
   velocities components (the position ones are zero) of the eigenvalue
    corresponding to the harmonic instability.}
  \label{fig:hardinsmode}
\end{figure}

Figure \ref{fig:Bquarhardsta01} shows that at $\varepsilon=0.01$
the tangent bands have moved upwards and, therefore, the stability
of the system is maintained. The system loses its stability at
$\varepsilon=0.022$ where a harmonic bifurcation appears. The
instability mode, shown in \ref{fig:hardinsmode} with its
multipliers, is an asymmetric extended one. Simulations done
perturbing with it the dark breather give rise to an small
oscillation with both sides of the chain out off phase,
superimposed to the dark breather one, but the {\em darkness} is
preserved. After that, there are Krein crunches due to the bands
mixing with the usual properties.

\begin{figure}
  \begin{center}
    \includegraphics[width=0.75\textwidth]{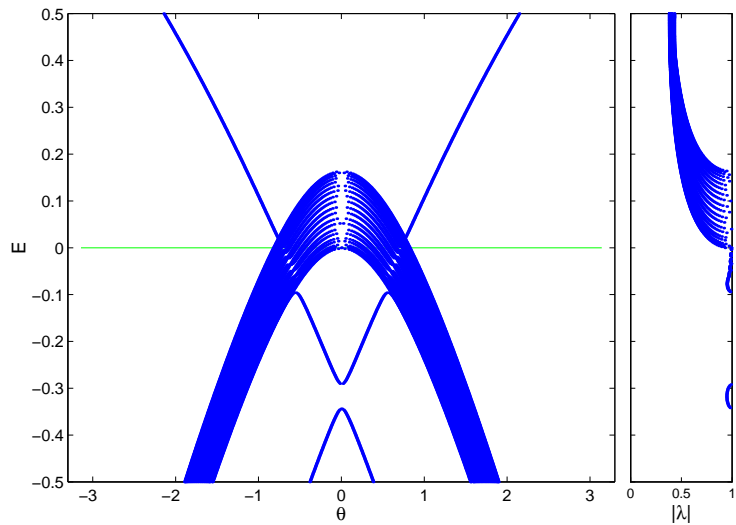}
  \end{center}
  \caption{Zoom of the band structure for a quartic hard on--site potential
   with attractive interaction at $\varepsilon=0.041$ and $\wb=1.2$.}
  \label{fig:Bquarhardsta041}
\end{figure}

\section{Finite size effects}
\label{sec:finite}

This final section is dedicated to comment some preliminary
results relative to the finite size effects on instability of dark
breathers. Important differences appear with respect to the case
of bright breathers \cite{MA98}.

In the case of bright breathers, there are two kinds of
size--dependent bifurcations. The origin of them relies on the
nature of the localization of the colliding eigenvalues. If the
colliding eigenvalues are extended, there appear ``instability
bubbles'', i.e. the Floquet eigenvalues abandon the unit circle
after the collision but return afterwards. These bifurcations
disappear when the system is infinite. Alternatively, a localized
eigenvalues can collide with a band of extended eigenvalues. In
this case, the instability bubbles also occur but they persist
even though the system is infinite.

In order to study the case of dark breathers, we have chosen a
Morse on--site potential with repulsive interaction. As a result
of this analysis, the collision of extended eigenvalues always
implies a cascade of subharmonic bifurcations (see section
\ref{sec:darkstability}) independently of the size of the system.
It is due to the fact that they are due to excited bands, whose
Floquet arguments belong to the first Brillouin zone, losing
intersections at $\theta=\pi$. It is different from the case of
bright breathers where this kind of instabilities are due to modes
corresponding to oscillators at rest and the bands must be reduced
to the first Brillouin zone.

\begin{figure}
\begin{tabular}{cc}
    \includegraphics[height=55mm]{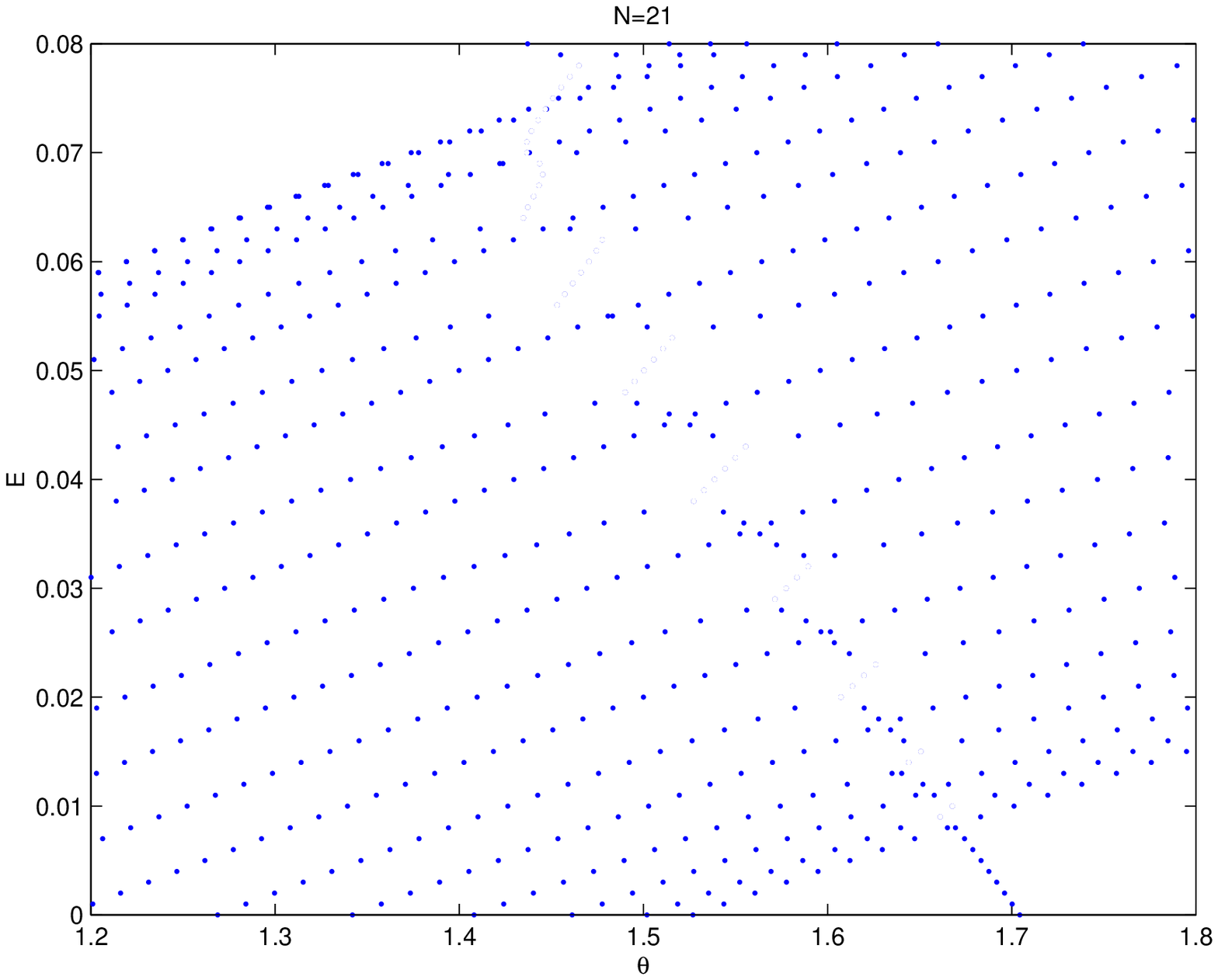} &
    \includegraphics[height=55mm]{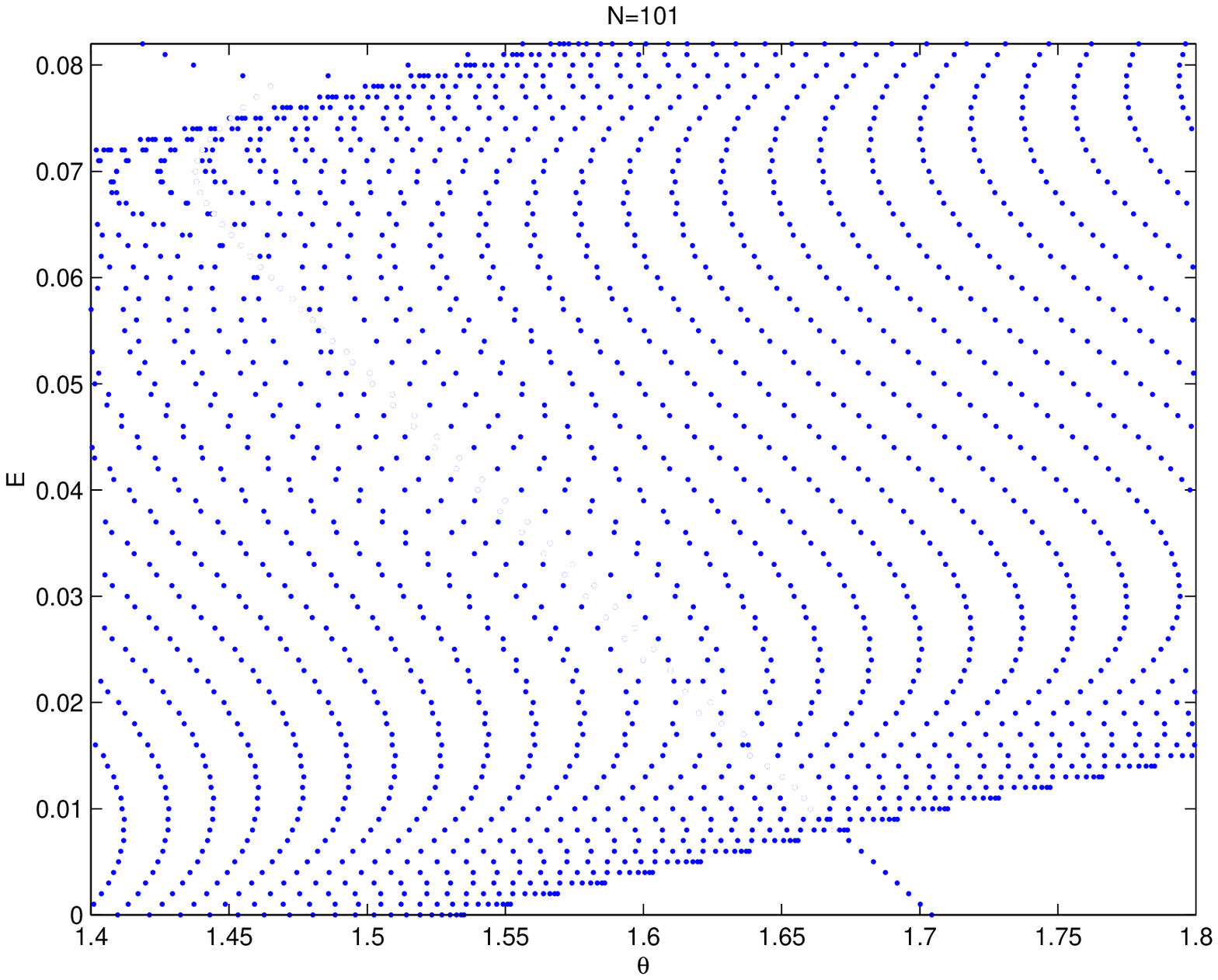}\
    \\
\end{tabular}
\caption{Zoom of the band scheme for N=21 and N=101 and a Morse
potential ($\ee=0.02$ and $\wb=0.8$). It can be observed the
appearance of wiggles when N=21, which are much wider when N=101
these wiggles are almost invisible because they occupy the width
of the band.} \label{fig:wiggles}
\end{figure}

\begin{figure}
\begin{tabular}{cc}
    \includegraphics[height=55mm]{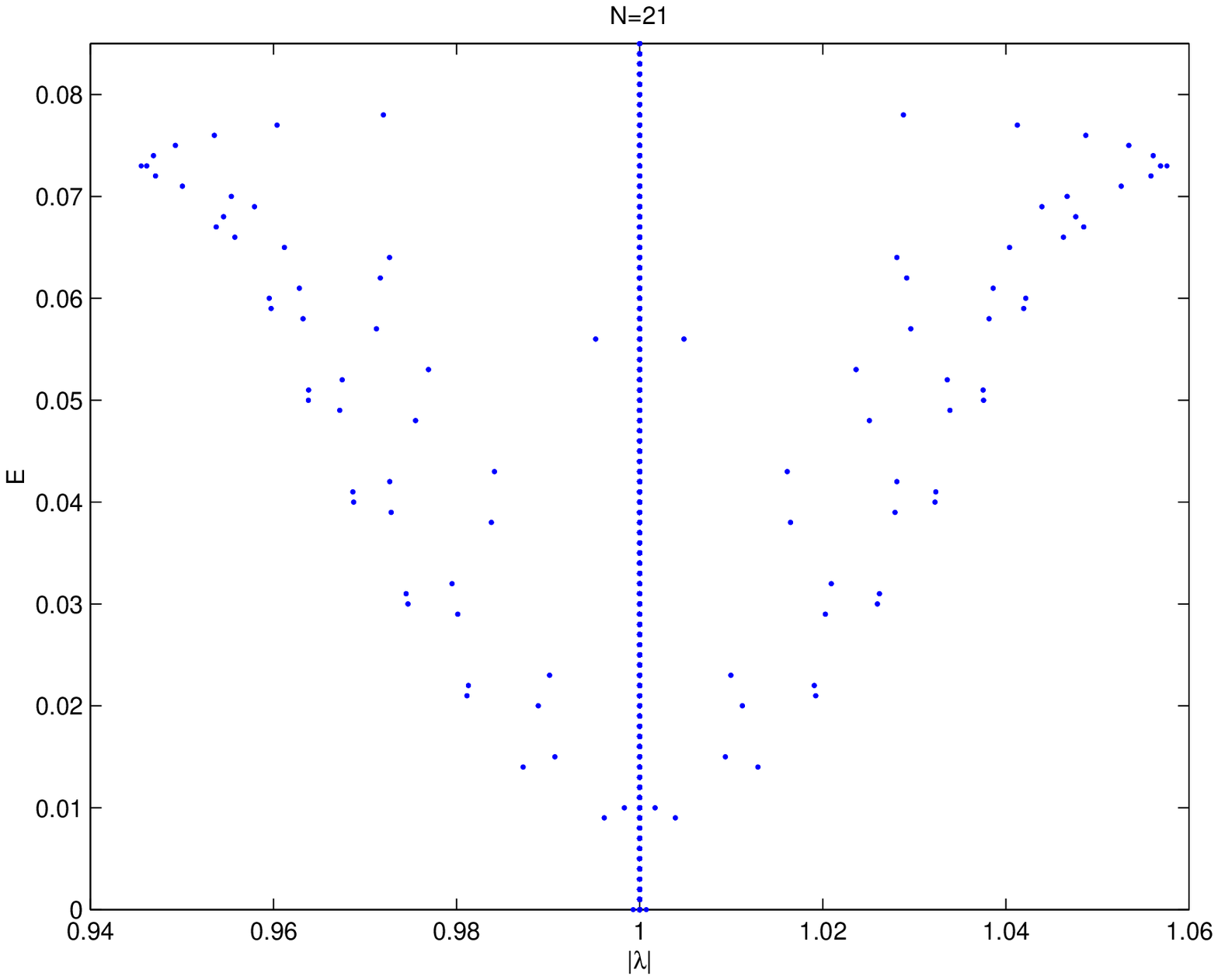} &
    \includegraphics[height=55mm]{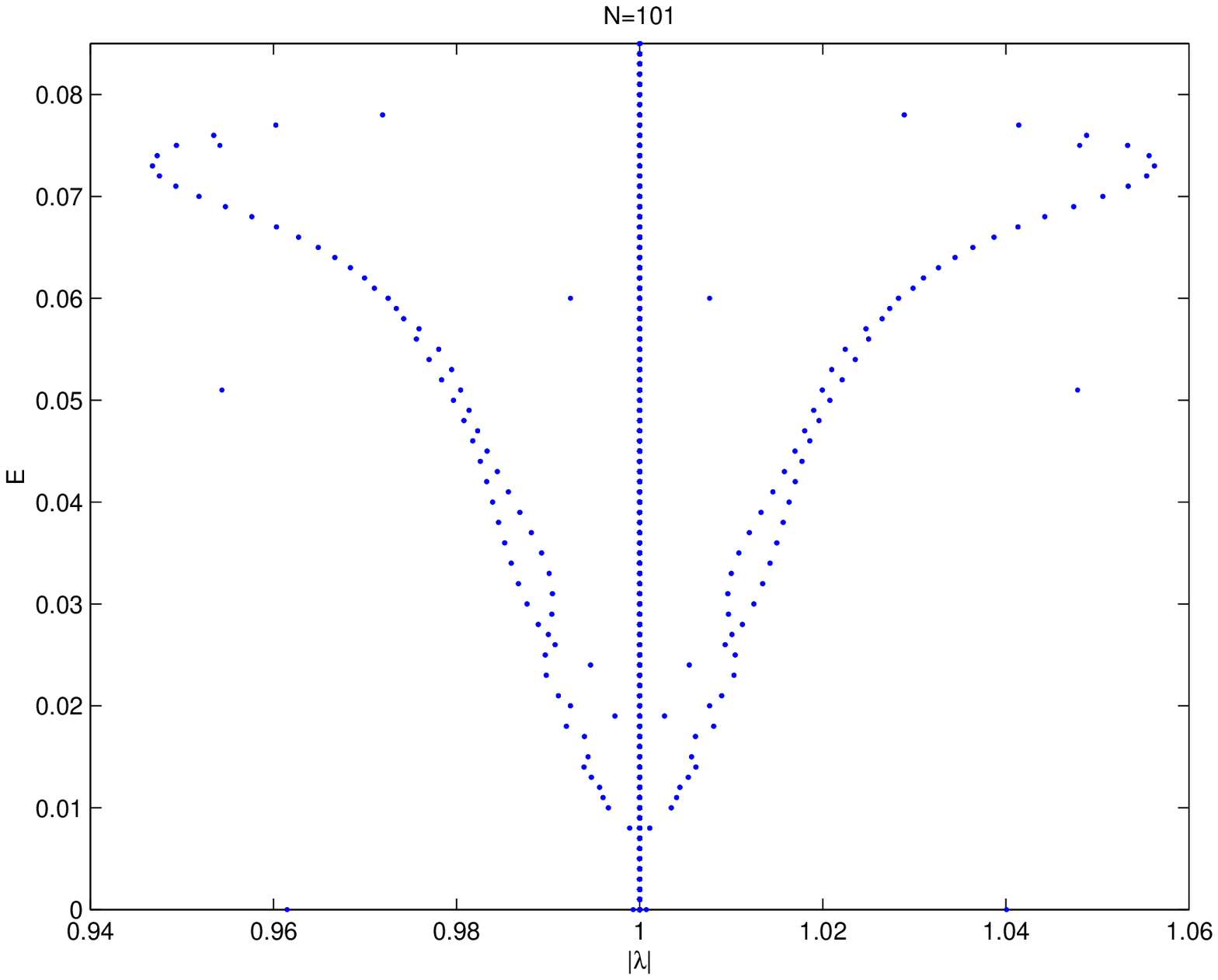}\
    \\
\end{tabular}
\caption{Moduli of the Floquet eigenvalues corresponding to the
Krein crunches for N=21 and N=101 and a Morse potential
($\ee=0.02$ and $\wb=0.8$). It can be observed the appearance of
instability bubbles when N=21. However, these bubbles transform
into an instability that persist up to a value of the coupling for
which the breather is unstable because of the subharmonic
bifurcations.} \label{fig:bubbles}
\end{figure}

Also, an eigenvalue corresponding to an extended mode can collide
with a localized mode. In this case, the rest (localized mode)
band and the excited (extended modes) bands mix (see Figures
\ref{fig:wiggles}). There appear ``wiggles'' that imply the
opening of gaps in the band scheme, which are the origin of the
instability bubbles. However, when the size of the system is
increased, the wiggles widen until they occupy almost the width of
the quasi-continuous extended mode bands. This fact implies that,
although the instability bubbles disappear, the Krein crunches are
unavoidable (see Figure \ref{fig:bubbles}). Nevertheless, the
breathers are robust despite of the existence of these
instabilities.

\section{Dark breathers stability in dissipative systems}
\label{sec:darkdisip}

There appear some important differences in the study of the
stability of dark breathers in dissipative systems with respect to
the Hamiltonian case.

In order to perform this study, we start from the dissipative
Frenkel--Kontorova model \cite{MFMF01}:

\begin{equation}
\ddot u_n+\gamma\dot u_n+
V'(u_n)+\varepsilon(2u_n-u_{n+1}-u_{n-1})=F\sin{\wb t}.
\end{equation}

where $\gamma$ is a damping parameter and $F$ the amplitude of an
external force.

\begin{figure}
\begin{tabular}{cc}
    \multicolumn{2}{c}{(a)}\\
    \multicolumn{2}{c}{
      \includegraphics[width=.45\textwidth]{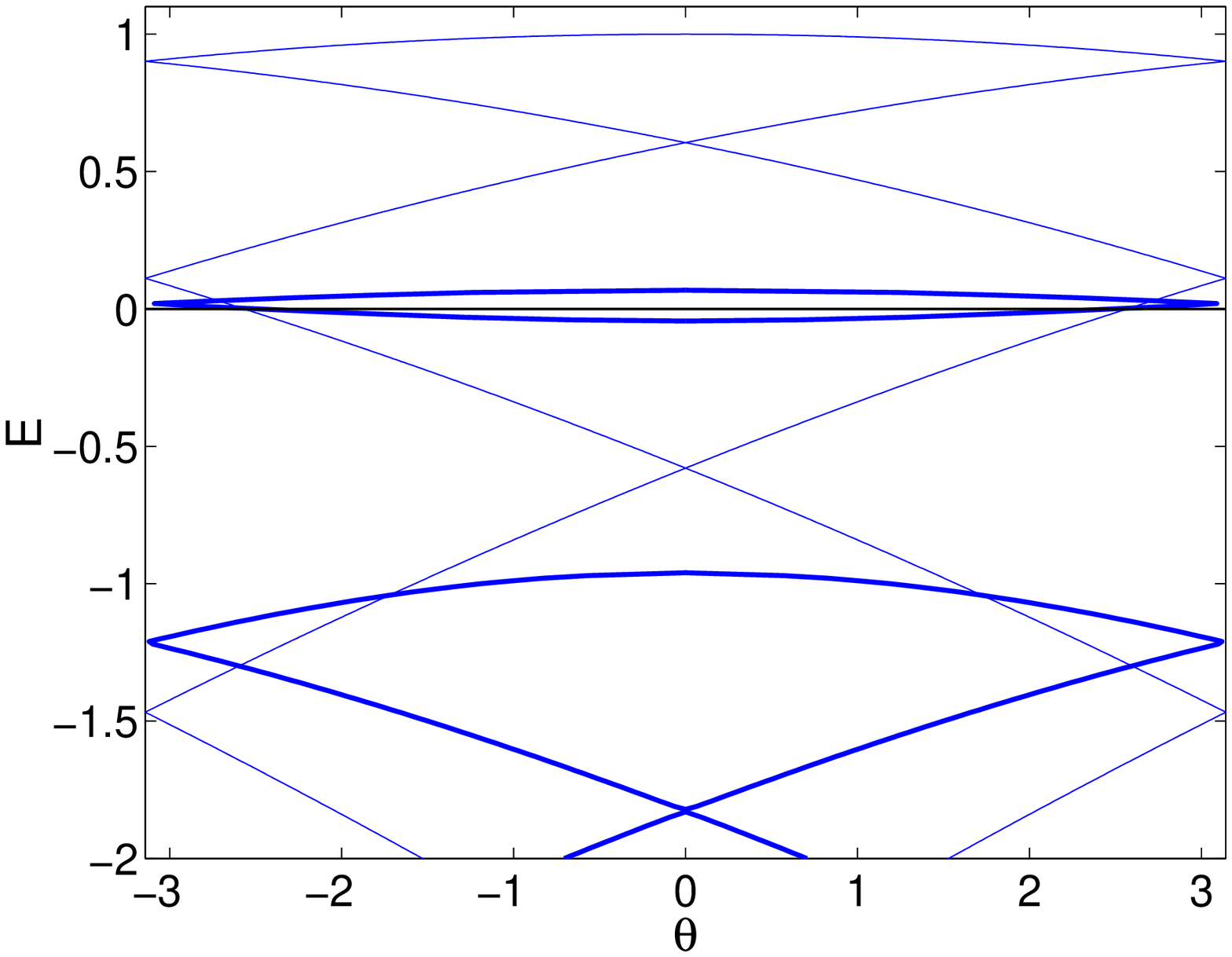}}\\
    (b) & (c) \\
    \includegraphics[width=.45\textwidth]{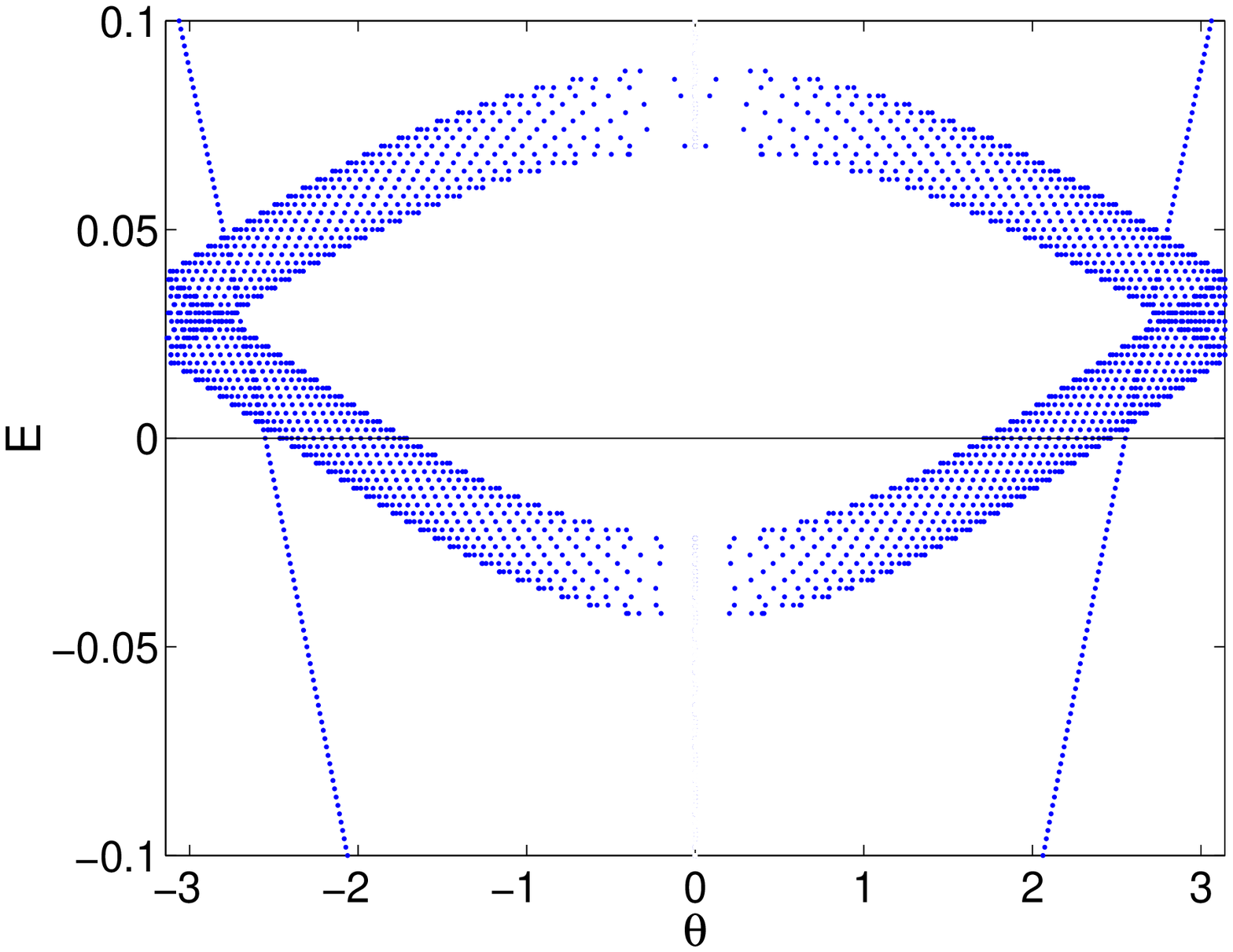} &
    \includegraphics[width=.45\textwidth]{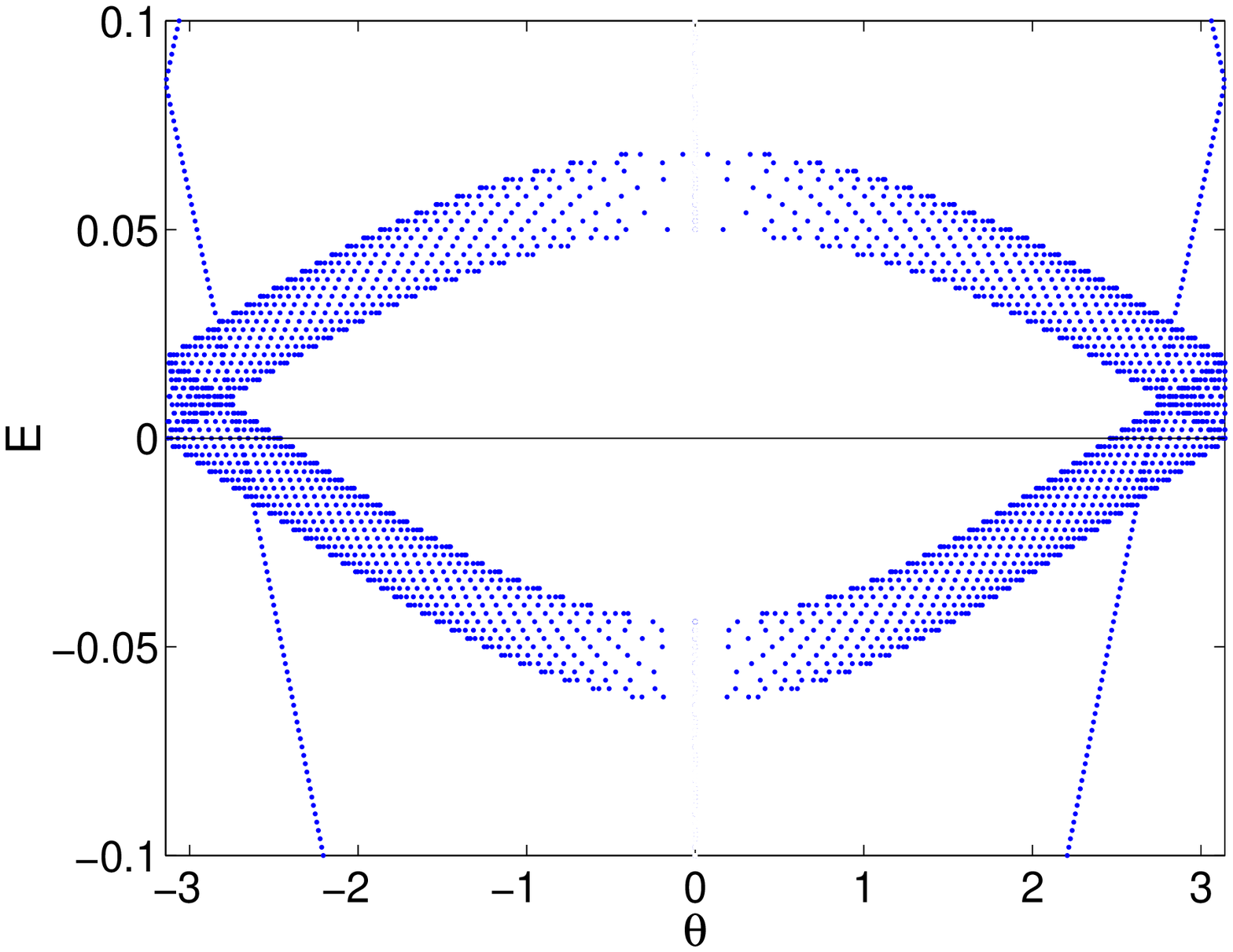}\\
\end{tabular}
\caption{Band diagrams in the anticontinuous limit (top) and for a
coupling $\varepsilon=\pm0.005$ (bottom) for a dark breather in
the Frenkel--Kontorova model. The parameters are: $\wb=0.2\pi$,
$\gamma=0.02$ y $F=0.02$. (b) corresponds to an attractive
interaction potential, whereas (c) corresponds to a repulsive
potential. The dark breather is stable in both cases.}
\label{banddis}
\end{figure}

In the anticontinuous limit, there are no band tangent to the
$E=0$-axis (figure \ref{banddis}a). It implies that, when the
coupling is introduced, there are no losing of intersections of
the bands with independence of the sign of the coupling constant
($\varepsilon$). Furthermore, the band corresponding to the
oscillator at rest do not mix with the band of the background and
oscillatory instabilities do not appear. These phenomena are shown
in figure \ref{banddis}.

\section{Conclusions}
\label{sec:conclusions}

In this paper we have explored the existence and stability of dark
breathers in one--dimensional Klein--Gordon models, for
frequencies far enough from the linear frequencies to  the DNLS
approximation be justified . We have found stable dark breathers
in several types of them. For systems with soft on--site
potential, there are no stable dark breathers if the coupling
between particles is attractive, but with a repulsive coupling the
stability is assured for fairly high values of the coupling
parameter $\varepsilon$. As $\varepsilon$ increases instability
bifurcations due to Krein crunches take place as a consequence of
the band mixing between the rest oscillator band and the dark
background. Eventually, the system experiences subharmonic
bifurcations at $-1$ that make the breather unstable. If the soft
on--site potential is symmetric the subharmonic bifurcations are
avoided and the final instability is caused by harmonic
bifurcations. For systems with hard on--site potentials, the
situation is reversed and the dark breathers are stable with
attractive coupling and unstable with repulsive one. They
experience harmonic bifurcations at $+1$ apart from the Krein
crunches. An analysis of larger systems shows that these
bifurcations persist even though the system is infinite.
Dissipative systems are, however, stable for both types of
coupling.

We are now on the project of performing a wider study of dark
breathers in dissipative systems. Another interesting aspect is to
study the relationship of the vibration pattern of dark breathers
with the hardness of the on--site potential and the type of
coupling. The ansatz $u_n\rightarrow (-1)^n\,u_n$, which
transforms repulsive coupling into attractive one and vice versa,
for symmetric potentials, suggests, and it has been shown for the
DNLS equations, and we also have checked numerically, that the
stability conditions are reversed are low coupling, but the
bifurcations for $|\wb -\omega_0|$ large enough, highly asymmetric
potentials as found in chemical and biological systems, and larger
coupling are worth studying.

\section*{Acknowledgments}

This work has been supported by the European Union under the RTN
project, LOCNET, HPRN--CT--1999--00163.

JFR Archilla acknowledges Informatics and Mathematical Modelling
at DTU for hospitality, while this work was started and Yu~B
Gaididei for valuable suggestions.

J Cuevas acknowledges an FPDI grant from `La Junta de
Andaluc\'{\i}a'.

\newcommand{\noopsort}[1]{} \newcommand{\printfirst}[2]{#1}
  \newcommand{\singleletter}[1]{#1} \newcommand{\switchargs}[2]{#2#1}
  \ifx\undefined\allcaps\def\allcaps#1{#1}\fi\ifx\undefined\allcaps\def\allcap%
s#1{#1}\fi

\end{document}